\global\def\draftcontrol{0}

   \def\versionno{exotic2}
\catcode`\@=11

\expandafter\ifx\csname draftcontrol\endcsname\relax\global\def\draftcontrol{0}
\fi

{\count255=\time\divide\count255 by 60
\xdef\hourmin{\number\count255}
\multiply\count255 by-60\advance\count255 by\time
\xdef\hourmin{\hourmin:\ifnum\count255<10 0\fi\the\count255}}
\def\draftdate{\number\month/\number\day/\number\year\ \ \ \hourmin }

\newcommand\makepapertitle{\par
  \begingroup
    \renewcommand\thefootnote{\@fnsymbol\c@footnote}%
    \def\@makefnmark{\rlap{\@textsuperscript{\normalfont\@thefnmark}}}%
    \long\def\@makefntext##1{\parindent 1em\noindent
            \hb@xt@1.8em{%
                \hss\@textsuperscript{\normalfont\@thefnmark}}##1}%
     \newpage
     \global\@topnum\z@   % Prevents figures from going at top of page.
     \@makepapertitle
     \thispagestyle{empty}\@thanks
  \endgroup
  \setcounter{footnote}{0}%
  \global\let\thanks\relax
  \global\let\makepapertitle\relax
  \global\let\@makepapertitle\relax
  \global\let\@thanks\@empty
  \global\let\@author\@empty
  \global\let\@date\@empty
  \global\let\@title\@empty
  \global\let\title\relax
  \global\let\author\relax
  \global\let\date\relax
  \global\let\and\relax
  \def\version{\let\version\@version\@gobble}
}
\def\@makepapertitle{%
  \newpage
   \ifnum\draftcontrol=1 {}
   \version\versionno
   \vskip 3em%
   \else
   \hfill\hbox to 3cm {\parbox{4cm}{\@pubnum}\hss}%
   \vskip 3em%
   \fi
   \begin{center}%
   \let \footnote \thanks
     {\LARGE {\@title}}%
     \vskip 1.5em%
     {\normalsize%\large
       \lineskip .5em%
       \begin{tabular}[t]{c}%
         \@author
       \end{tabular}\par}%
     \vskip 1.5em%
     {\@bstract}%
     \end{center}%
     \vskip 1.5em
     \@date%
   \par
}

\gdef\@pubnum{}
\def\pubnum#1{%
  \gdef\@pubnum{#1}}

\gdef\@bstract{}
\def\Abstract#1{%
  \gdef\@bstract{%
   \parbox{\textwidth-0pc}{%
   \centerline{\bf Abstract}\penalty1000%
\kern.2cm%
\noindent%\abstractfont \baselineskip=12pt
\renewcommand\baselinestretch{1.0}%
{#1}}}
}

\def\ps@paper{\let\@mkboth\@gobbletwo%
     \ifnum\draftcontrol=1
    \def\@oddfoot{\hbox to \textwidth{\tiny \versionno \hfil\tiny\draftdate}%
    \hskip -\textwidth \hbox to \textwidth{\hfil\rm\thepage\hfil}}%
     \else\def\@oddfoot{\hbox to \textwidth{\hfil\rm\thepage\hfil}}
     \fi
     \let\@evenfoot\@oddfoot
}
\def\body{\clearpage
          \pagestyle{paper}
    }
\def\@version#1{\ifnum\draftcontrol=1
\typeout{}\typeout{#1}\typeout{}
\vskip3mm\centerline{\hbox{\fbox{\normalsize{\tt DRAFT -- #1 -- }
                   {\draftdate}}}}\vskip3mm
\fi}
\let\version\@version
\long\def\eqlabel#1{\ifnum\draftcontrol=1
                    \tag@false  % there are some problems with multline without this
                    \tag*{(\theequation) \hbox to -0.2cm{\hspace{0cm}\small{#1}\hss}}
                    \refstepcounter{equation}
                    \edef\@currentlabel{\theequation}
                    \ltx@label{#1}          % use old LaTeX \label instead of new definition
                    \else
                    \label{#1}
                    \fi
                    }
\let\st@bibitem\@bibitem
\let\st@lbibitem\@lbibitem
\ifnum\draftcontrol=1
  \def\@bibitem#1{%
    \st@bibitem{#1}\a@@label{#1}\ignorespaces}
  \def\@lbibitem[#1]#2{%
    \st@lbibitem[#1]{#2}\a@@label{#2}\ignorespaces}
  \def\a@@label#1{%
    \gdef\a@lab{\smash{\normalfont\small#1}}
    \ifvmode
      \if@inlabel
        \global\setbox\@labels\hbox{%
          \llap{\a@lab\let\a@lab\relax
                \kern\@totalleftmargin\kern\marginparsep}%
          \box\@labels}%
      \fi
    \fi}
\fi
\documentclass[12pt,letterpaper]{article}
\usepackage{amsmath,amssymb,array,calc,epsfig,rotating,bm}
\usepackage[sort]{cite}
\usepackage{graphicx}
\usepackage{psfrag,verbatim}
\usepackage{xcolor}

\ifnum\draftcontrol=1
\fi
\renewcommand\baselinestretch{1.25}
\renewcommand\section{\@startsection {section}{1}{\z@}%
                                   {-3.5ex \@plus -1ex \@minus -.2ex}%
                                   {2.3ex \@plus.2ex}%
                                   {\normalfont\large\bfseries}}
\renewcommand\subsection{\@startsection{subsection}{2}{\z@}%
                                   {-3.25ex\@plus -1ex \@minus -.2ex}%
                                   {1.5ex \@plus .2ex}%
                                   {\normalfont\normalsize\bfseries}}
\renewcommand\subsubsection{\@startsection{subsubsection}{3}{\z@}%
                                   {-3.25ex\@plus -1ex \@minus -.2ex}%
                                   {1.5ex \@plus .2ex}%
                                   {\normalfont\normalsize\it}}
\renewcommand\paragraph{\@startsection{paragraph}{4}{\z@}%
                                   {-3.25ex\@plus -1ex \@minus -.2ex}%
                                   {1.5ex \@plus .2ex}%
                                   {\normalfont\normalsize\bf}}

\def\revise#1       {\raisebox{-0em}{\rule{3pt}{1em}}%
                     \marginpar{\raisebox{.5em}{\vrule width3pt\
                     \vrule width0pt height 0pt depth0.5em
                     \hbox to 0cm{\hspace{0cm}{%
                     \parbox[t]{4em}{\raggedright\footnotesize{#1}}}\hss}}}}

\newcommand\nxt[1]  {\\\fnxt#1}
\newcommand{\ie}{{\it i.e.,}\ }

\def\calc         {{\cal C}}

\def\cale         {{\cal E}}
\def\calf         {{\cal F}}

\def\call         {{\cal L}}

\def\calo         {{\cal O}}

\def\reals        {{\mathbb R}}
\def\zet          {{\mathbb Z}}

\def\del          {\partial}

\def\qft          {{{\rm QFT}_3}}
\def\cft         {{{\rm CFT}_3}}

\def\sqr#1#2{{\vcenter{\vbox{\hrule height.#2pt
 \hbox{\vrule width.#2pt height#1pt \kern#1pt
 \vrule width.#2pt}\hrule height.#2pt}}}}

\newcommand{\ft}[2]{{\textstyle{\frac{#1}{#2}}}}

\def\aa1{\phi}
\def\cc1{\psi}

\catcode`\@=12

\begin{document}

%%%
%%%%%% text starts here
%%%%%%%%%

\title{\bf Thermal order in holographic CFTs and no-hair theorem violation in black branes}

\date{May 15, 2020}
%\date\today

\author{
Alex Buchel \\
\it Department of Applied Mathematics\\
\it Department of Physics and Astronomy\\ 
\it University of Western Ontario\\
\it London, Ontario N6A 5B7, Canada\\
\it Perimeter Institute for Theoretical Physics\\
\it Waterloo, Ontario N2J 2W9, Canada\\[0.4cm]
}

\Abstract{We present a large class of holographic models where the boundary
${\mathbb R}^{2,1}$ dimensional conformal field theory has a thermal phase
with a spontaneously broken global symmetry. The dual black branes in a
Poincare patch of asymptotically $AdS_4$ violate the no-hair theorem.
}

\makepapertitle

\body

\version\versionno
%\tableofcontents

In \cite{Chai:2020zgq} the authors asked an interesting question whether an ``order''
(a thermal phase with a spontaneously broken global symmetry) is always lost at
high temperatures\footnote{As in \cite{Chai:2020zgq},
we consider equilibrium phases of the theories without a chemical potential
for the conserved global $U(1)$ symmetries.}? There is a number of holographic models
known where the answer is ``yes'',
\ie there is a critical temperature $T_{crit}$, such that for $T>T_{crit}$
there is a phase with a spontaneously broken global discrete, $\zet_2$,  symmetry
\cite{Buchel:2009ge,Bosch:2017ccw,Buchel:2017map} or a continuous, $U(1)$,
symmetry \cite{Buchel:2018bzp}. The holographic models mentioned correspond to
boundary quantum field theories with a mass scale: a coupling of a relevant operator in the
model \cite{Buchel:2009ge} or a strong coupling scale of the Klebanov-Strassler
 gauge theory \cite{Klebanov:2000hb} in \cite{Buchel:2018bzp}. Ultimately,
 it is this mass  scale that determines $T_{crit}$.

But is it possible to have an exotic phenomenon of the global symmetry breaking in the
ultraviolet (as first suggested in \cite{Buchel:2009ge}) in a conformal theory?
There is no scale to determine $T_{crit}$, and the thermal symmetric and the symmetry broken
phases must exist for all temperatures. Necessarily, the {\it distinct}
holographic duals to these phases
--- the black branes with the translationary invariant horizons --- 
would violate the no-hair theorem. The purpose of this note is to present
explicit examples of such holographic models.
An impatient reader can simply jump to the discussion of the
holographic models in {\bf Step3}. Rather, we follow the construction
route from massive QFTs.

$\bullet$ {\bf Step1}\footnote{This is a review of \cite{Buchel:2009ge}.}. Consider
the effective four-dimensional gravitational bulk
action\footnote{We set the radius of an asymptotic $AdS_4$ 
geometry to unity.} ,
dual to a $\qft$ on $\reals^{2,1}$,
\begin{equation}
\begin{split}
S_\qft=&S_{\cft}+S_{r}+S_i=\frac{1}{2\kappa^2}\int dx^4\sqrt{-\gamma}\left[\call_{\cft}+\call_{r}+\call_i\right]\,,
\end{split}
\eqlabel{s4}
\end{equation}
\begin{equation}
\call_{\cft}=R+6\,,\qquad \call_r=-\frac 12 \left(\nabla\phi\right)^2+\phi^2\,,\qquad 
\call_i=-\frac 12 \left(\nabla\chi\right)^2-2\chi^2-g \phi^2 \chi^2 \,,
\eqlabel{lc}
\end{equation}
where we split the action into   a conformal part $S_{\cft}$; its deformation by a relevant
operator $\calo_r$; and a sector $S_i$ involving an irrelevant operator $\calo_i$ 
along with its mixing with $\calo_r$ under the renormalization group flow, specified by a constant $g$. In all our numerical analysis we set $g=-100$. The precise value of $g$ does not matter, as long as $g$ is sufficiently
negative, see \cite{Buchel:2009ge}.
The four dimensional gravitational constant $\kappa$ is related to a central charge $c$ of the 
UV fixed point as 
\begin{equation}
c=\frac{192}{ \kappa^2}\,.
\eqlabel{cg}
\end{equation}
We assume the scaling dimension of $\calo_r$ to be $\Delta_r=2$.
The scaling dimension of $\calo_i$ is $\Delta_i=4$ . Generically, we turn on the non-normalizable
coefficient of $\phi$, corresponding, the nonzero coupling $\Lambda$  of the dual
operator $\calo_r$. The effective action \eqref{s4} has a $\zet_2\times \zet_2$ discrete symmetry that acts as a
parity transformation on 
the scalar fields $\phi$ and $\chi$. The discrete symmetry $\phi\leftrightarrow -\phi$ is explicitly broken by a relevant deformation 
of the $\cft$; while the $\chi\leftrightarrow -\chi$ symmetry is broken spontaneously. 

The thermal states of the $\qft$ are dual to  the  black brane solutions in \eqref{s4} with translationary invariant 
horizons:
\begin{equation}
ds_4^2=\frac{\alpha^2 a(x)^2}{(2x-x^2)^{2/3}}\biggl(-(1-x)^2 dt^2+\left[dx_1^2+dx_2^2\right]\biggr)+g_{xx}\ dx^2\,,\ \phi=\phi(x)\,,\ 
\chi=\chi(x)\,,
\eqlabel{background}
\end{equation}
where the radial coordinate $x\in (0,1)$. The constant $\alpha$ is an arbitrary scale parameter, and the metric warp factor $g_{xx}$
is determined algebraically from $a,\phi,\chi$. Solving the equations of motion from  \eqref{s4} and \eqref{lc} with the background 
ansatz \eqref{background}, we find, 
\begin{equation}
\begin{split}
a=&1-\frac{1}{40}\ p_1^2\ x^{2/3}-\frac{1}{18}\ p_1 p_2\ x+\calo(x^{4/3})\,,\\
\phi=&p_1\ x^{1/3}+p_2\ x^{2/3}+\frac{3}{20} p_1^3 x+\calo(x^{4/3})\,,\\
\chi=&\chi_4\left(x^{4/3}+\left(\frac 17 g -\frac{3}{70}\right)p_1^2\ x^2+\calo(x^{7/3})\right)\,,
\end{split}
\eqlabel{boundary}
\end{equation}
near the $AdS_4$ boundary $x\to 0_+$, and 
\begin{equation}
\begin{split}
a=a_0^{h}+a_1^h\ y^2+\calo(y^4)\,,\qquad \phi=p_0^h+\calo(y^2)\,,\qquad  \chi=c_0^h+\calo(y^2)\,,
\end{split}
\eqlabel{hor}
\end{equation}
near the black brane horizon $y=1-x\to 0_+$. 
Apart from the overall scaling factor $\alpha$, a black brane solution is specified with the three UV coefficients
$\{p_1,p_2,\chi_4\}$ and the four  IR coefficients  $\{a_0^h,a_1^h,p_0^h,c_0^h\}$. The UV  parameter $p_1$ determines
the coupling constant of the relevant operator $\calo_r$ as
\begin{equation}
\Lambda\equiv p_1\alpha\,,
\eqlabel{deflambda}
\end{equation}
while the remaining parameters determine the Hawking temperature $T$ of the black brane, its entropy density $s$,
the energy density $\cale$, and the free energy density $\calf$  as follows:
\begin{equation}
\begin{split}
&\left(\frac{8\pi T}{\Lambda}\right)^2=\frac{6 (a^h_0)^3 (6-2 (c^h_0)^2+(p^h_0)^2-g (p^h_0)^2 (c^h_0)^2)}{p_1^2(3 a^h_1+a^h_0)}\,,\qquad
\frac{\hat{s}}{\Lambda^2}\equiv \frac{384}{c}\ \frac{s}{\Lambda^2}= \frac{4\pi (a_0^h)^2}{p_1^2}\,,\\
&\frac{\hat{\cale}}{\Lambda^3}\equiv\frac{384}{c}\ \frac{\cale}{\Lambda^3}=\frac{1}{p_1^3} \left(2-\frac 16  p_1 p_2\right)\,,\qquad
\frac{\hat{\calf}}{\Lambda^3}\equiv\frac{384}{c}\ \frac{\calf}{\Lambda^3}=\frac{384}{c}\ \frac{\cale- Ts}{\Lambda^3}\,.
\end{split}
\eqlabel{therel}
\end{equation}
Additionally, the thermal expectation values of the operators $\calo_r$ and $\calo_i$ are given by
\begin{equation}
\langle\calo_r\rangle\propto \langle\hat{\calo_r}\rangle=p_2\,,\qquad \langle\calo_i\rangle\propto \langle\hat{\calo_i}\rangle=\chi_4\,.
\eqlabel{vevs}
\end{equation}
The latter expectation value, \ie $\chi_4$, serves as an order parameter for the spontaneous
breaking of the global $\zet_2$ symmetry.

\begin{figure}[t]
\begin{center}
\psfrag{f}[cc][][0.8][0]{$\hat{\calf}/(\pi T)^3$}
\psfrag{e}[cc][][0.8][0]{$\cale/\calf$}
\psfrag{l}[cc][][0.8][0]{${\Lambda}/{8\pi T}$}
\includegraphics[width=3in]{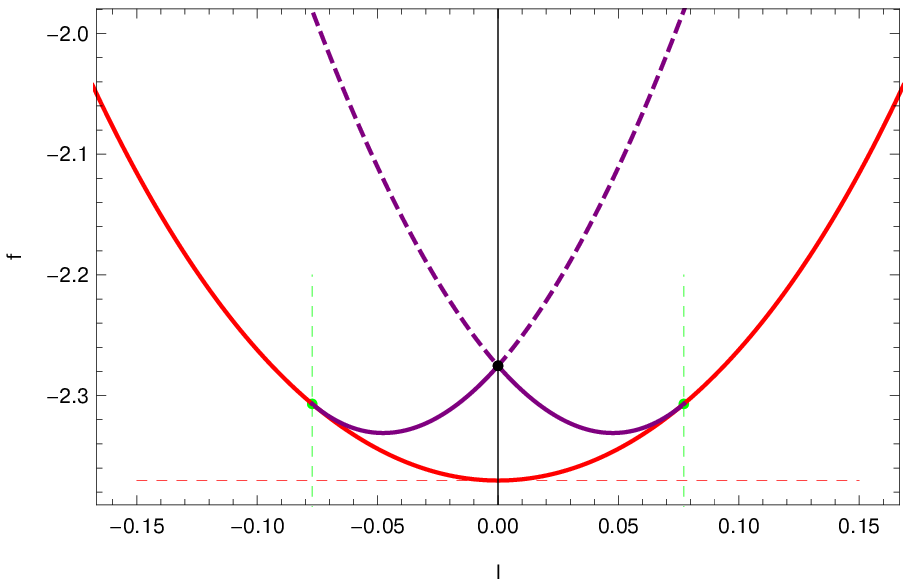}\,
\includegraphics[width=3in]{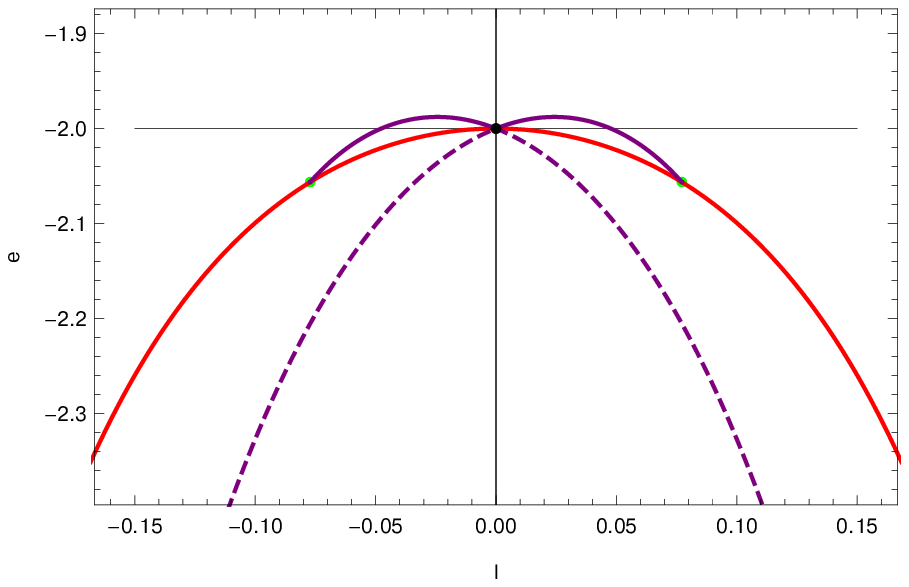}
\end{center}
  \caption{Thermal phases of the  $\qft$ with a dual holographic gravitational action \eqref{s4}.
  The left panel shows the free energy density $\hat{\calf}$, see \eqref{therel}, and the right panel shows $\cale/\calf$ as a function of $\Lambda/8\pi T$.
  At $\Lambda=0$ the theory has $\zet_2\times \zet_2$ global symmetry. This symmetry is broken
  to $\zet_2$ at $\Lambda\ne 0$ --- represented by the red curves. The purple curves
  represent phases with the spontaneously broken $\zet_2$ symmetry. They connect with the
  second order phase transition (green dots, vertical dashed green lines) to a symmetric
  phase at $T_{crit}\propto \Lambda$, see  \eqref{tcrit}. The symmetry broken
  phases (solid purple curves) exist for $T\ge T_{crit}$ all the way to the infinite temperature.
 The infinite temperature phase --- represented by a black dot --- is our first example of the
 $\cft$ with spontaneously broken global symmetry, see \eqref{black}.  The symmetry breaking
 phases can be smoothly extended past the infinite temperature (alternatively across $\Lambda=0$)
 --- denoted by the purple dashed
 curves.
} \label{fig1}
\end{figure}

\begin{figure}[t]
\begin{center}
\psfrag{c}[cc][][0.8][0]{$\langle\hat{\calo_i}\rangle$}
\psfrag{p}[cc][][0.8][0]{$\langle\hat{\calo_r}\rangle$}
\psfrag{l}[cc][][0.8][0]{${\Lambda}/{8\pi T}$}
\includegraphics[width=3in]{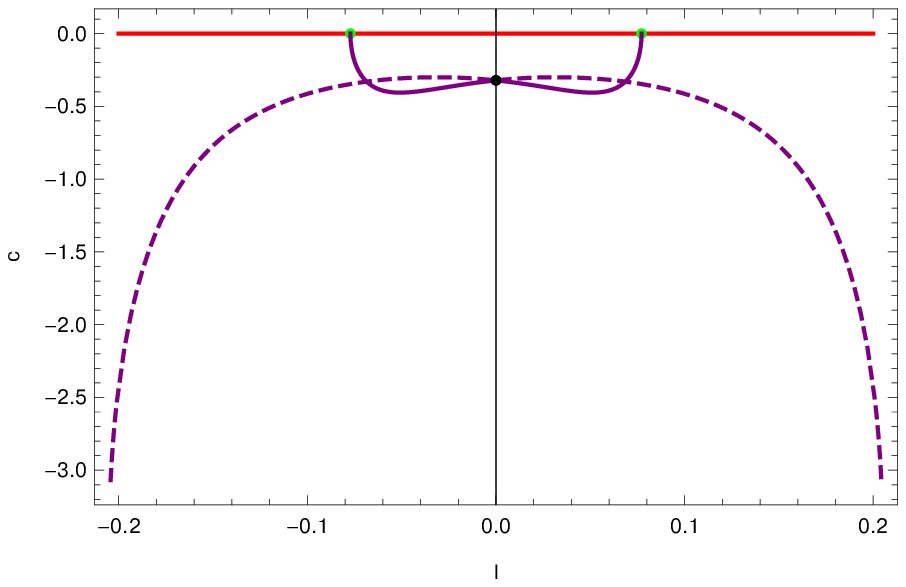}\,
\includegraphics[width=3in]{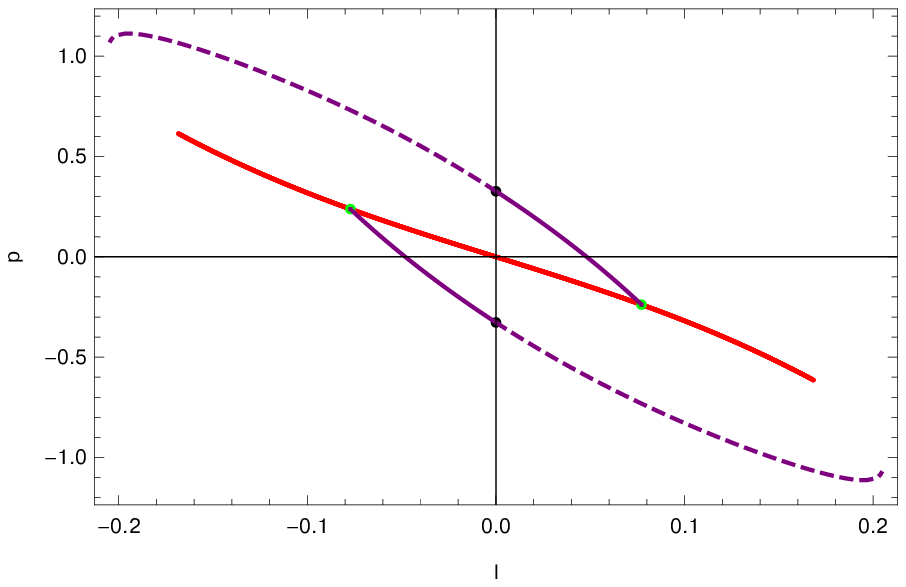}
\end{center}
  \caption{The left panel: the  order parameter $\langle\calo_i\rangle$ for
the spontaneous breaking of the global $\zet_2$ symmetry in the $\qft$.
The right panel: the thermal expectation value of $\langle\calo_r\rangle$
in the  $\qft$. The color coding is the same
as in fig.~\ref{fig1}.
} \label{fig2}
\end{figure}

\begin{figure}[t]
\begin{center}
\psfrag{c}[cc][][1][0]{$c_s^2$}
\psfrag{t}[cc][][1][0]{$T/{|\Lambda|}$}
\includegraphics[width=3in]{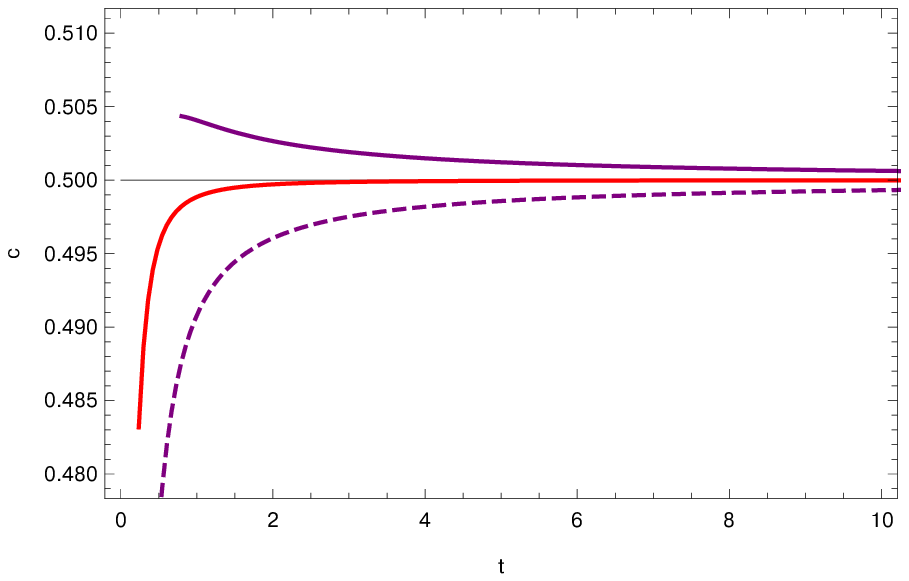}
\end{center}
  \caption{The speed of the sound waves in various equilibrium phases of the $\qft$ plasma.
  The color coding is the same
as in fig.~\ref{fig1}.
} \label{fig2a}
\end{figure}

We briefly review the numerical procedure used to collect the data for the thermal phases of the
model \eqref{s4}, reported in figs.~\ref{fig1}-\ref{fig2a}, see \cite{Buchel:2009ge} for
additional details:
\nxt The red curves in figs.~\ref{fig1}-\ref{fig2a} represent the $\zet_2$ symmetric thermal
phase. Here, the bulk scalar $\chi(x)\equiv 0$, correspondingly with $\chi_4=0$ in the UV,
see \eqref{boundary}, and $c_0^h=0$ in the IR, see \eqref{hor}.  We use the
shooting method first employed in \cite{Aharony:2007vg} to produce data sets
$\{p_2,a_0^h,a_1^h,p_0^h\}$ from varying $p_1$, related  to $\frac{T}{\Lambda}$, see \eqref{therel}.
Specifically, given $p_1$, we numerically solve the pair of
the coupled second-order ODEs for $a$ and $\phi$ from the UV for $x\in(0,\ft 12]$ and IR
for $y=(0,\ft 12]$, using the corresponding
asymptotic expansions \eqref{boundary} and \eqref{hor}, and adjust $\{p_2,a_0^h,a_1^h,p_0^h\}$
to ensure the continuity of $a,\phi$ and their first derivatives at $x=y=\ft 12$.
From the collected data we compute the thermodynamics of the symmetric phase using
\eqref{therel}. The speed of the sound waves in  plasma,
see fig.~\ref{fig2a}, is obtained numerically evaluating the derivative 
\begin{equation}
c_s^2=-\frac{\del\calf}{\del\cale}=-\frac{\frac{d}{dp_1}\frac{\hat\calf}{\Lambda^3}}
{\frac{d}{dp_1}\frac{\hat\cale}{\Lambda^3}}\,.
\eqlabel{defcs}
\end{equation}
\nxt  The purple curves in figs.~\ref{fig1}-\ref{fig2a} represent the thermal
phases of the model \eqref{s4} with the spontaneous breaking of the $\zet_2$ symmetry, \ie
with $\chi_4\ne 0$. Here, we need to solve three coupled second-order ODEs for $\{a,\phi,\chi\}$.
Given $p_1$, the asymptotic expansions \eqref{boundary} and \eqref{hor} are characterized
by $\{p_2,\chi_4,a_0^h,a_1^h,p_0^h,c_0^h\}$. Once again we employ the shooting method
of \cite{Buchel:2009ge}. In practice, to ensure that we obtain solutions with $\chi_4\ne 0$,
we fix initially $\chi_4=-1$ and obtain from the shooting algorithm
the corresponding set $\{p_1,p_2,a_0^h,a_1^h,p_0^h,c_0^h\}$. Additional data sets are obtained
from this {\it initial} set varying $p_1$ (in small increments), as for the symmetric phase.
A remarkable fact, that was not appreciated in \cite{Buchel:2009ge}, is that we can vary $p_1$
in the symmetry broken phase through $p_1=0$ (represented by a black dot in
figs.~\ref{fig1} and \ref{fig2}). According to \eqref{therel}, see the expression for
$\frac{T}{\Lambda}$, this corresponds to
reaching an infinitely high temperature and {\it extending past the infinite temperature}
(at fixed $\Lambda$), or alternatively crossing $\Lambda=0$, \ie the conformal point.
Turns out that there are two distinct thermal phases with the spontaneously broken $\zet_2$ symmetry:
we use the solid purple curves to denote the phase with  the speed of the
sound waves $c_s^2>\ft 12$, and  the dashed purple curves to denote the phase with  the speed of the
sound waves $c_s^2<\ft 12$, see fig.~\ref{fig2a}.

A selection\footnote{There is a tower of the symmetry broken thermal
phases  similar to those discussed here \cite{Buchel:2009ge}.}
of thermal phases of the $\qft$ is presents in fig.~\ref{fig1}.
Notice that the horizontal axes correspond to $\Lambda/(8\pi T)$, so the region close to the
origin corresponds to high temperatures, \ie $T\gg \Lambda$. We show\footnote{In all cases we verified that the first law of thermodynamics $0=d\cale/(Tds)-1\bigg|_{\Lambda={\rm const}}$
is true numerically
to $\sim 10^{-10}$ or better.} the $\zet_2$ symmetric phases
with $\langle\hat{\calo_i}\rangle=0$
(red curves) and the symmetry broken phases $\langle\hat{\calo_i}\rangle\ne 0$ (purple curves).
The left panel shows the reduced free energy density $\hat{\calf}$, see \eqref{therel}.
Note that for the $AdS_4$-Schwarzschild black brane
\begin{equation}
\frac{\hat{\calf}}{(\pi T)^3}\bigg|_{{red},\Lambda=0}=-\frac{64}{27}\,,
\eqlabel{ads4s}
\end{equation}
denoted by a red dashed horizontal line. The solid purple curves connect to the red curve
with the second-order transition, green dots, at
\begin{equation}
T_{crit}=0.515597\ |\Lambda|:\qquad \lim_{T\to T_{crit} } \langle\calo_i\rangle =0\,,
\eqlabel{tcrit}
\end{equation}
denoted by the green dashed vertical lines.
As discovered in \cite{Buchel:2009ge}, the symmetry broken phases exist only in the UV, \ie for
$T\ge T_{crit}$. They extend to infinitely high temperatures, denoted by the black dot. 
The black dot is our first example of the $\cft$ with the spontaneously broken global
discrete symmetry, in this particular case $\zet_2\times \zet_2$:
\begin{equation}
\cft:\qquad  \frac{\hat{\calf}}{(\pi T)^3}=-2.275317\,,\qquad
\langle\hat{\calo_r}\rangle=\pm 0.326946\,,\qquad \langle\hat{\calo_i}\rangle=\pm 0.321581\,,
\eqlabel{black}
\end{equation}
where the uncorrelated $\pm$ signs represent the 4-fold degeneracy of the symmetry
broken phases.

In any thermal phase, symmetric or symmetry broken,
the equation of state of a $\cft$ is $\cale=-2 \calf$. 
In the right panel of fig.~\ref{fig1} we plot $\cale/\calf$ in the $\qft$ as a function of $\Lambda/8\pi T$:
both the red curve and the purple curves pass through $(-2)$ in the UV
(with a numerical accuracy of $\sim 10^{-11}$).

The symmetry broken phases (solid purple curves) can be smoothly extended ``past the infinite
temperature'' --- the dashed purples curves.  We have been able to reliably construct the
dashed purple phases only for $|\Lambda|/(8\pi T)\ \lesssim\ 0.2043$.

In fig.~\ref{fig2} we present the order parameter $\langle\hat{\calo_i}\rangle$ for
the spontaneous breaking of the global $\zet_2$ symmetry (the left panel),
and the thermal expectation value of $\langle\hat{\calo_r}\rangle$ (the right panel).
The color coding is the same as
in fig.~\ref{fig1}. Notice that as one lowers the temperature along the
dashed purple curves, the absolute value of the order parameter sharply increases
as  $|\Lambda|/(8\pi T)\ \to\   0.2043$. This is the main reason why we could not extend
these phases to low temperatures\footnote{We expect that the corresponding black branes
have a singular horizon in this limit.}.

In fig.~\ref{fig2a} we present the speed of the sound waves in various thermal phases of the
$\qft$ plasma.  The color coding is the same
as in fig.~\ref{fig1}. All the equilibrium phases, symmetric and with the spontaneously
broken $\zet_2$ symmetry, are thermodynamically and dynamically stable \cite{Buchel:2005nt}.
The speed of the sound waves approach a conformal value in the limit $T/\Lambda\to \infty$.

An example of a $\cft$ with a spontaneously broken $\zet_2\times \zet_2$ --- the black dot phase in
fig.~\ref{fig1} --- has a larger free energy density than that of the symmetric phase, dual to
 the $AdS_4$ -Schwarzschild black brane:
\begin{equation}
\frac{\hat{\calf}}{(\pi T)^3}\bigg|_{black\ dot}\qquad >\qquad \frac{\hat{\calf}}{(\pi T)^3}\bigg|_{{red},\Lambda=0}\,.
\eqlabel{compare}
\end{equation}
Thus, this phase is metastable\footnote{It would be extremely interesting to
understand the dynamics of the first order phase transition, in particular, to compute the
wall tension of the symmetric phase bubble in the symmetry broken thermal phase.}.
In what follows, we ask the question whether we can introduce an additional
``knob'' in a holographic model
\eqref{s4} to potentially reverse the inequality \eqref{compare}, and have a symmetry
broken phase to dominate in a canonical ensemble\footnote{No examples of such
models known.}. This leads up to {\bf Step2}.

 $\bullet$ {\bf Step2}. Consider a smooth constant $b$ parameter deformation of the model
 \eqref{s4}:
 \begin{equation}
 \begin{split}
 S_\qft\ \to\ S_{\qft^b}\qquad \Longleftrightarrow\qquad \call_i\to\ \call_i^b\equiv
 -\frac 12 \left(\nabla\chi\right)^2-2\chi^2-g \phi^2 \chi^2-b \chi^4\,,
\end{split}
\eqlabel{bdef}
 \end{equation}
with the remaining parts of the effective action left unchanged. 
Note that the ${\qft^{b}}$ has the same global symmetries as the $\qft$.
Additionally, the second-order phase transitions associated with the
spontaneous breaking of $\zet_2$ symmetry ($\chi\leftrightarrow -\chi$)
happens at $T_{crit}$ given by \eqref{tcrit}, independent of the deformation parameter
$b$.

It is straightforward to repeat the analysis of the thermal phase diagram of the model.
In the thermodynamic relations \eqref{therel} there is only one
change\footnote{Of course, all the  UV and IR parameters
$\{p_2,\chi_4,a_0^h,a_1^h,p_0^h,c_0^h\}$ will develop an implicit $b$ dependence.}:
\begin{equation}
\left(\frac{8\pi T}{\Lambda}\right)^2=\frac{6 (a^h_0)^3 (6-2 (c^h_0)^2+(p^h_0)^2-g (p^h_0)^2 (c^h_0)^2-b (c_0^h)^4)}{p_1^2(3 a^h_1+a^h_0)}\,.
\eqlabel{therelb}
\end{equation}

\begin{figure}[t]
\begin{center}
\psfrag{d}[cc][][1][0]{$(\hat{\calf}^b-\hat{\calf}^{0})/(\pi T)^3$}
\psfrag{l}[cc][][1][0]{${\Lambda}/{8\pi T}$}
\includegraphics[width=3in]{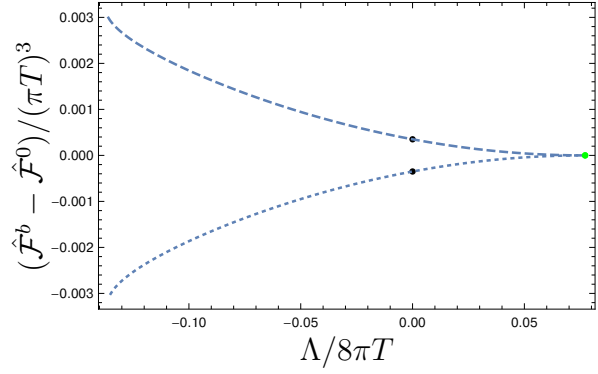}
\end{center}
  \caption{The reduced free energy densities $\hat{\calf}^b$  of the $\zet_2$ symmetry broken
  thermal phases of the $\qft^b$ model, see \eqref{bdef}, compare to the
  free energy density $\hat{\calf}^0$  of the corresponding  phases
  of the $\qft$ model, see \eqref{s4}. The dashed curve corresponds to $b=1$ and the
  dotted curve corresponds to  $b=-1$ deformation parameter. The green dot indicates
  the $b$-independent critical temperature $T_{crit}$, see \eqref{tcrit}. The black dots
  represent new examples of $\cft^b$ with spontaneously broken $\zet_2\times \zet_2$
  global symmetry. 
 } \label{fig3}
\end{figure}

In fig.~\ref{fig3} we show the effect of the deformation parameter
$b$ on the reduced free energy density $\hat{\calf}^b$  of the symmetry
broken phase in the $\qft^b$ model,
compare to the corresponding free energy density in the $\qft$ model, denoted by $\hat{\calf}^0$
(this is the solid magenta curve in the left panel of fig.~\ref{fig1}).  
The dashed curve corresponds to $b=1$ and the dotted curve corresponds to $b=-1$.
They originate from the same green dot, representing the critical temperature $T_{crit}$, as in
\eqref{tcrit}. Following the symmetry broken phases in the $\qft^b$ model to infinitely
high temperature, we identify conformal phases, $\cft^b$,  with the spontaneously broken
$\zet_2\times \zet_2$ global symmetry --- these are the two black dots in fig.~\ref{fig3}. 
Since we want to reduced the free energy density as much as possible, we now study $\cft^b$ models
for $b<0$\footnote{Notice that the gravitational scalar potential in \eqref{bdef}
is unbounded for $b<0$.
This does not immediately signal any pathologies --- many scalar potentials in
top-down supersymmetric holographic models are unbounded. See \cite{Buchel:2017map}
for further discussion.}.

\begin{figure}[t]
\begin{center}
\psfrag{f}[cc][][0.8][0]{$\hat{\calf}^b/(\pi T)^3$}
\psfrag{b}[cc][][0.8][0]{$b$}
\psfrag{p}[cc][][0.8][0]{$\langle\hat{\calo_r}\rangle$}
\includegraphics[width=3in]{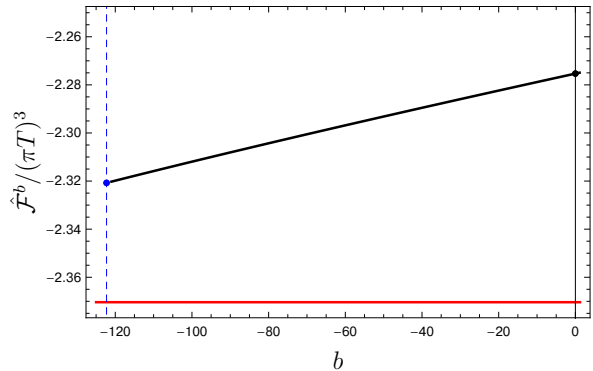}
\end{center}
  \caption{The reduced free energy density $\hat{\calf}^b$ in the $\cft^b$ model
  as a function of the deformation parameter $b$ (the solid black line).
  The black dot represents the same symmetry broken conformal phase
  as the black dot in fig.~\ref{fig1}. The $\cft^b$ symmetry broken phase
  exists only for $b\ge b_{crit}$, see \eqref{bcrit}, denoted by a vertical dashed blue line ---
  it terminates with a blue dot $\cft^\chi$, see \eqref{cftpsi}. The horizontal
  red  line represents the reduced free energy density of the
  $AdS_4$-Schwarzschild black brane.
} \label{fig5}
\end{figure}

In fig.~\ref{fig5} we follow (the solid black curve)
the free energy density in the $\cft^b$ model from $b=0$ (represented by the black dot ---
the same black dot as in fig.~\ref{fig1}) for $b<0$. The horizontal red line
is the value of the free energy density for the $AdS_4$-Schwarzschild black brane,
see \eqref{ads4s}. We find that the $\cft^b$ model allows for a symmetry broken phase
only for
\begin{equation}
b\ge b_{crit}= -122.272\,.
\eqlabel{bcrit}
\end{equation}
The critical value of the deformation parameter is denoted by  a vertical
dashed blue line.

\begin{figure}[t]
\begin{center}
\psfrag{b}[cc][][0.8][0]{$b$}
\psfrag{p}[cc][][0.8][0]{$\langle\hat{\calo_r}\rangle$}
\psfrag{c}[cc][][0.8][0]{$\langle\hat{\calo_i}\rangle$}
\includegraphics[width=3in]{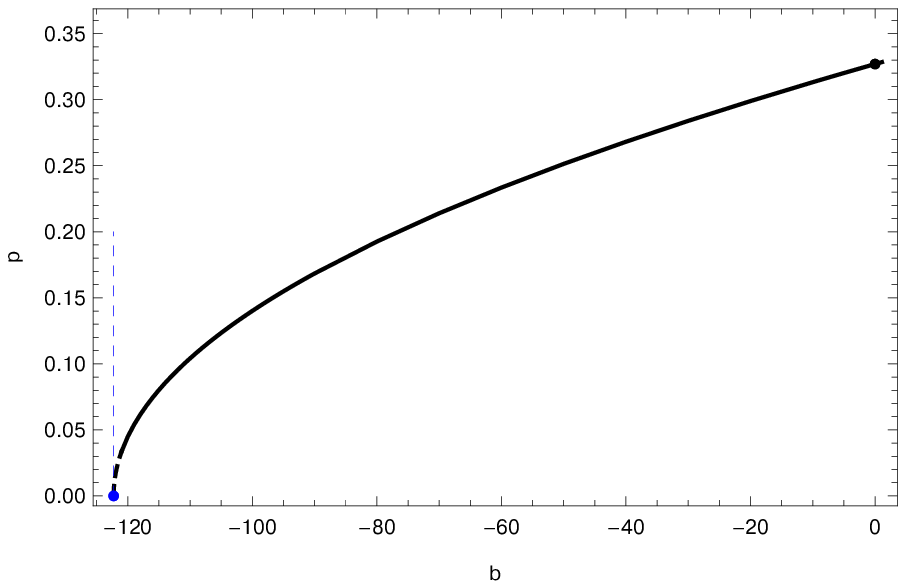}\,
\includegraphics[width=3in]{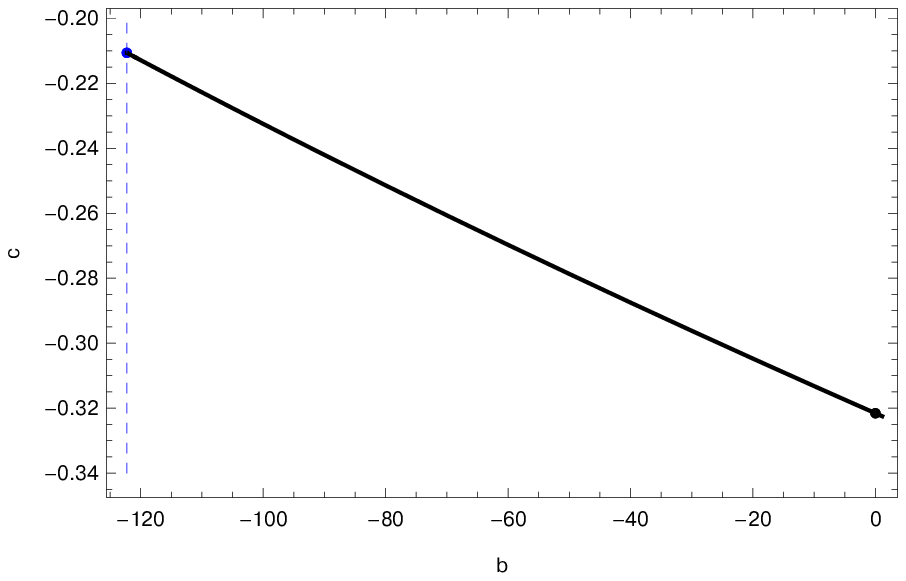}
\end{center}
  \caption{The symmetry breaking order parameters
$\langle\hat{\calo_r}\rangle$ and $\langle\hat{\calo_i}\rangle$ in the $\cft^b$ model
as a function of $b$. The $\langle\hat{\calo_r}\rangle$ order parameter vanishes
as $b\to b_{crit}$, denoted by the vertical dashed blue lines.
However, the $\langle\hat{\calo_i}\rangle$ order parameter remains finite at $b=b_{crit}$
(the blue dot). 
}\label{fig6}
\end{figure}

\begin{figure}[ht]
\begin{center}
\psfrag{f}[cc][][0.8][0]{$\hat{\calf}/(\pi T)^3$}
\psfrag{b}[cc][][0.8][0]{$b$}
\includegraphics[width=3in]{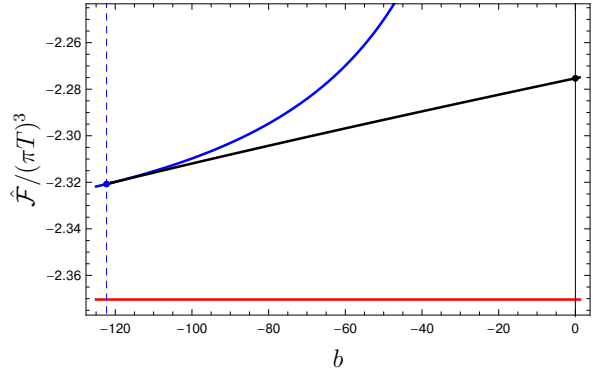}
\end{center}
  \caption{Same as in fig.~\ref{fig5}, with the addition of the symmetry breaking
  phase in the $\cft^\chi$ model, see \eqref{spsi} --- the solid blue curve.
} \label{fig7}
\end{figure}

\begin{figure}[ht]
\begin{center}
\psfrag{f}[cc][][0.8][0]{$\hat{\calf^\chi}/(\pi T)^3$}
\psfrag{b}[cc][][0.8][0]{$b$}
\includegraphics[width=2.5in]{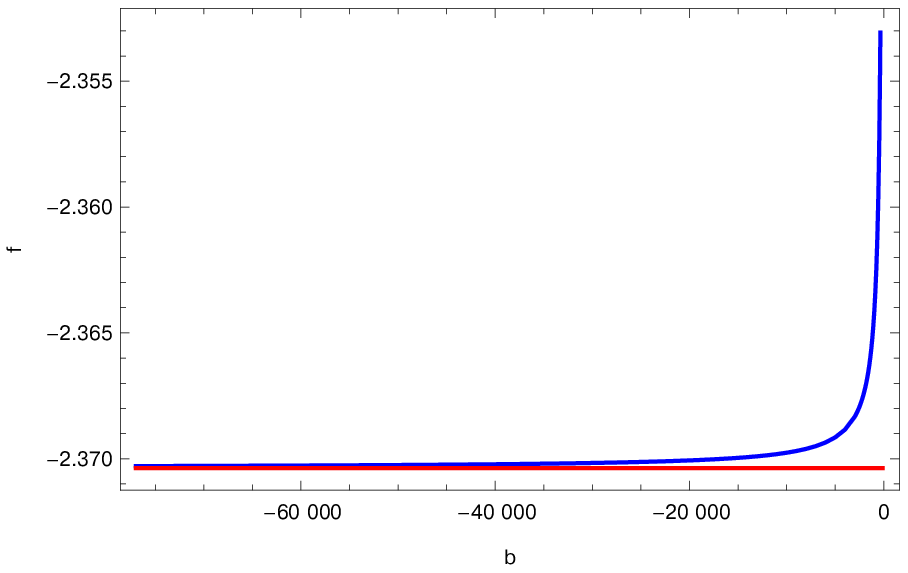}\,,
\includegraphics[width=2.5in]{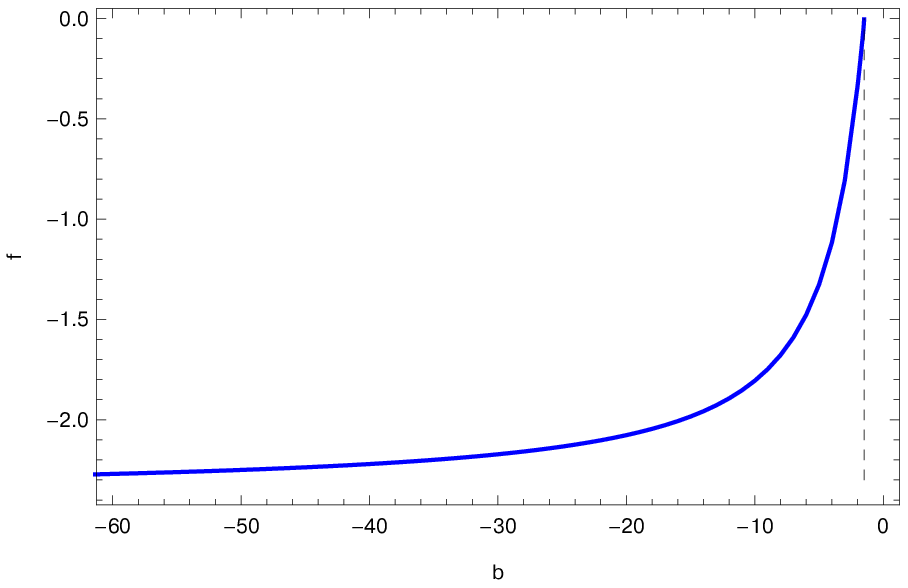}
\end{center}
  \caption{The reduced free energy density $\hat{\calf^\chi}$
  of the $\zet_2$ symmetry broken phase as a function of $b$.
This phase exists only for $b< -\frac 32$, represented by a vertical dashed black line in
the right panel. The horizontal
  red line represents the reduced free energy density of the
  $AdS_4$-Schwarzschild black brane. 
} \label{fig8}
\end{figure}

\begin{figure}[ht]
\begin{center}
\psfrag{i}[cc][][0.8][0]{$|\langle\hat{\calo_i}\rangle|^{-1}$}
\psfrag{b}[cc][][0.8][0]{$b$}
\includegraphics[width=2.5in]{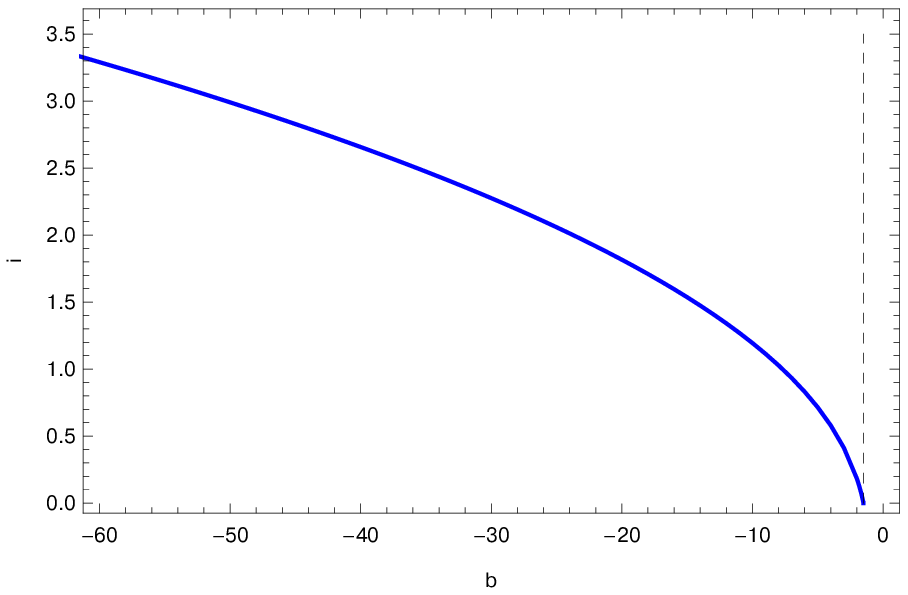}\,,
\includegraphics[width=2.5in]{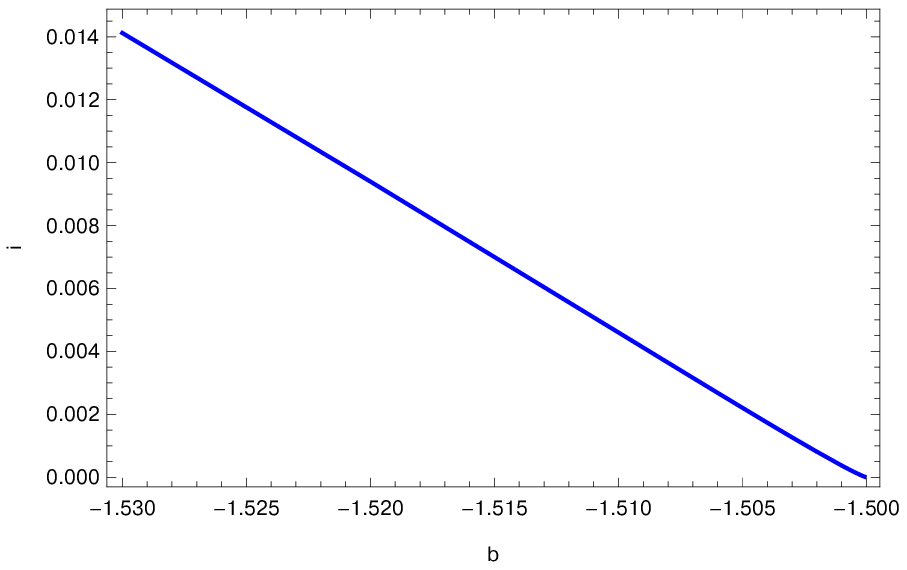}
\end{center}
  \caption{
The inverse of the order parameter $|\langle\hat{\calo_i}\rangle|$ for the
symmetry breaking in the $\cft^\chi$ model as $b\to -\frac 32_-$.
} \label{fig9}
\end{figure}

\begin{figure}[ht]
\begin{center}
\psfrag{d}[cc][][0.8][0]{$\ln\left[{\hat{\calf}^\chi}/{(\pi T)^3}+64/27\right]$}
\psfrag{z}[cc][][0.8][0]{$\ln |\langle\hat{\calo_i}\rangle|$}
\psfrag{l}[cc][][0.8][0]{$\ln(-b)$}
\includegraphics[width=2.5in]{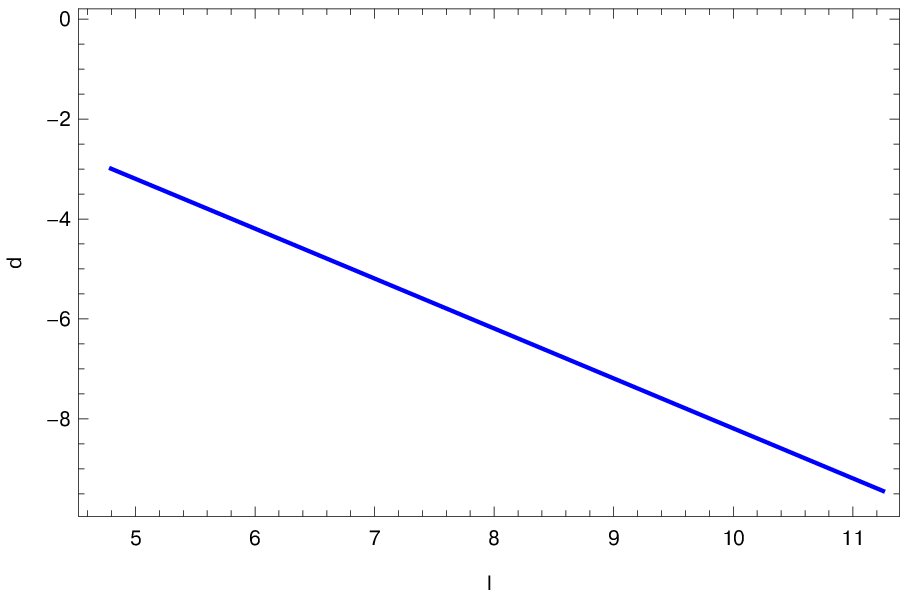}\,,
\includegraphics[width=2.5in]{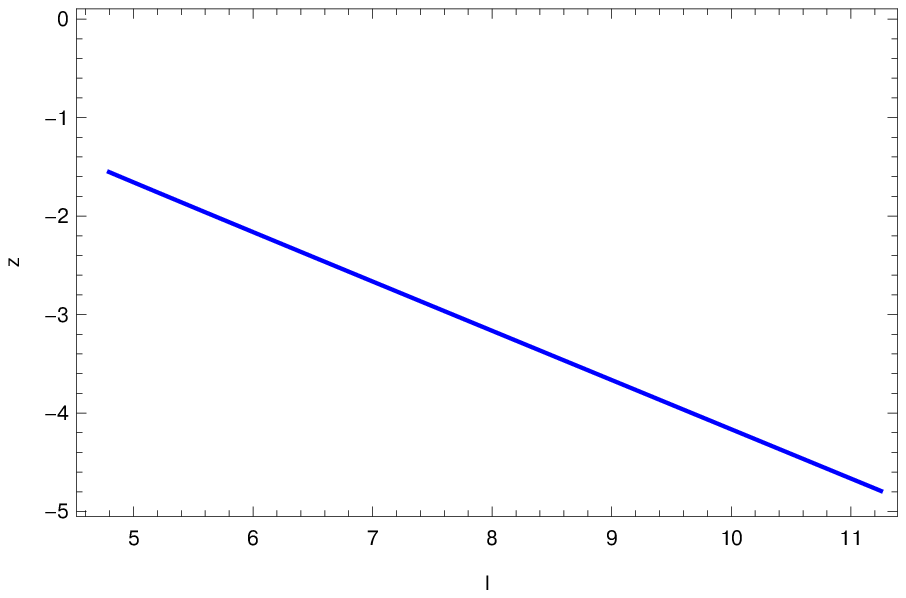}
\end{center}
  \caption{The reduced free energy density of the symmetry broken phase in the $\cft^\chi$
model always exceeds that of the symmetric phase. It approaches the latter as
$\propto \exp b$ in the limit $(-b)\to\infty$, see the left panel. In the right panel
we show the $\propto (-b)^{-1/2}$ scaling of the symmetry breaking order parameter 
in the $\cft^\chi$ model in the limit $(-b)\to \infty$.} \label{fig10}
\end{figure}

The origin of $b_{crit}$ is easy to understand as we follow the symmetry breaking order parameters
$\langle\hat{\calo_r}\rangle$ and $\langle\hat{\calo_i}\rangle$ in the $\cft^b$ model,
see fig.~\ref{fig6}. We find that while $\langle\hat{\calo_i}\rangle$ remains finite
in the limit $b\to b_{crit}$,
\begin{equation}
\langle\hat{\calo_r}\rangle\ \propto\ \left(b-b_{crit}\right)^{1/2}\,.
\eqlabel{relorder}
\end{equation}
Thus, precisely at $b=b_{crit}$ the black brane dual to the $\cft^b$ symmetry broken
phase "losses'' the $\phi$-hair, while maintaining the non-vanishing $\chi$-hair.
We call this "terminal'' conformal model $\cft^\chi$,
\begin{equation}
\begin{split}
&\cft^\chi\bigg|_{b=b_{crit}}\qquad \equiv\qquad  \lim_{b\to b_{crit}}\ \cft^b\,,\\
&\cft^\chi\bigg|_{b=b_{crit}}:\qquad  \frac{\hat{\calf}^\chi}{(\pi T)^3}\bigg|_{b=b_{crit}}=-2.32074\,,\qquad
\langle\hat{\calo_i}\rangle\bigg|_{b=b_{crit}}=\pm 0.21057\,,
\end{split}
\eqlabel{cftpsi}
\end{equation}
where again the $\pm$ signs indicate the
2-fold degeneracy due to the spontaneously broken $\zet_2$ 
global symmetry.

Once again,
\begin{equation}
\frac{\hat{\calf}^\chi}{(\pi T)^3}\bigg|_{b=b_{crit}}\qquad >\qquad
\frac{\hat{\calf}}{(\pi T)^3}\bigg|_{AdS_4-{\rm Schwarzschild}}\,.
\eqlabel{compare2}
\end{equation}
Can we do better with relaxing $b=b_{crit}$ constraint directly in the $\cft^\chi$ model?
This leads us to {\bf Step3}.

$\bullet$ {\bf Step3.} Consider a gravitational dual to $\cft^\chi$ model,
\begin{equation}
\begin{split}
S_{\cft^\chi}=&\frac{1}{2\kappa^2}\int dx^4\sqrt{-\gamma}\left[R+6
-\frac 12 \left(\nabla\chi\right)^2-2\chi^2-b\chi^4\right]\,,
\end{split}
\eqlabel{spsi}
\end{equation}
Clearly, the  $S_{\cft^\chi}$ model is a consistent truncation of the $S_{\qft^b}$ model
with $\phi\equiv 0$. The $\cft^\chi$ model has only $\zet_2$ global symmetry.
The symmetry broken blue dot phase in fig.~\ref{fig5} must also be a solution 
of the effective action \eqref{spsi}, except that now we are not restricted to
keep $b=b_{crit}$. In fig.~\ref{fig7} the solid blue curve traces  the free energy density 
of the symmetry broken phase in the $\cft^\chi$ model as a function of the parameter  $b$
in \eqref{spsi}.

The thermal phase with the spontaneously broken $\zet_2$ symmetry exists in the $\cft^\chi$
model for $b\le -\frac 32$, see fig.~\ref{fig8}.

As $b\to -\frac 32_-$, the order parameter for the symmetry breaking diverges as, see
fig.~\ref{fig9},
\begin{equation}
|\langle\hat{\calo_i}\rangle|\ \propto\ \frac{1}{-3/2-b}\,,\qquad {\rm as}\qquad
b\to -\frac 32_-\,.
\eqlabel{blimit}
\end{equation}

Notice that 
\begin{equation}
\frac{\hat{\calf}^\chi}{(\pi T)^3}\qquad > \qquad
\frac{\hat{\calf}}{(\pi T)^3}\bigg|_{AdS_4-{\rm Schwarzschild}}\,,
\eqlabel{compare3}
\end{equation}
for all value of $b$, see fig~\ref{fig10}. We find that as $b\to -\infty$,
\begin{equation}
\frac{\hat{\calf}^\chi}{(\pi T)^3}=\frac{\hat{\calf}}{(\pi T)^3}\bigg|_{AdS_4-{\rm Schwarzschild}}
+{\exp\calc}\ \left(\frac{1}{\sqrt{-b}}\right)^2\,,\qquad |\langle\hat{\calo_i}\rangle|\ \propto \frac{1}{\sqrt{-b}}\,,
\eqlabel{largeb}
\end{equation}
where $\calc\approx 1.81$.

\section*{Acknowledgments}
Research at Perimeter
Institute is supported by the Government of Canada through Industry
Canada and by the Province of Ontario through the Ministry of
Research \& Innovation. This work was further supported by
NSERC through the Discovery Grants program.\bibliographystyle{JHEP}
\bibliography{exotic2}

\providecommand{\href}[2]{#2}\begingroup\raggedright\begin{thebibliography}{1}

\bibitem{Chai:2020zgq}
N.~Chai, S.~Chaudhuri, C.~Choi, Z.~Komargodski, E.~Rabinovici and M.~Smolkin,
  \emph{{Thermal Order in Conformal Theories}},
  \href{https://arxiv.org/abs/2005.03676}{{\tt 2005.03676}}.

\bibitem{Buchel:2009ge}
A.~Buchel and C.~Pagnutti, \emph{{Exotic Hairy Black Holes}},
  \href{http://dx.doi.org/10.1016/j.nuclphysb.2009.08.017}{\emph{Nucl. Phys. B}
  {\bf 824} (2010) 85--94}, [\href{https://arxiv.org/abs/0904.1716}{{\tt
  0904.1716}}].

\bibitem{Bosch:2017ccw}
P.~Bosch, A.~Buchel and L.~Lehner, \emph{{Unstable horizons and singularity
  development in holography}},
  \href{http://dx.doi.org/10.1007/JHEP07(2017)135}{\emph{JHEP} {\bf 07} (2017)
  135}, [\href{https://arxiv.org/abs/1704.05454}{{\tt 1704.05454}}].

\bibitem{Buchel:2017map}
A.~Buchel, \emph{{Singularity development and supersymmetry in holography}},
  \href{http://dx.doi.org/10.1007/JHEP08(2017)134}{\emph{JHEP} {\bf 08} (2017)
  134}, [\href{https://arxiv.org/abs/1705.08560}{{\tt 1705.08560}}].

\bibitem{Buchel:2018bzp}
A.~Buchel, \emph{{Klebanov-Strassler black hole}},
  \href{http://dx.doi.org/10.1007/JHEP01(2019)207}{\emph{JHEP} {\bf 01} (2019)
  207}, [\href{https://arxiv.org/abs/1809.08484}{{\tt 1809.08484}}].

\bibitem{Klebanov:2000hb}
I.~R. Klebanov and M.~J. Strassler, \emph{{Supergravity and a confining gauge
  theory: Duality cascades and chi SB resolution of naked singularities}},
  \href{http://dx.doi.org/10.1088/1126-6708/2000/08/052}{\emph{JHEP} {\bf 08}
  (2000) 052}, [\href{https://arxiv.org/abs/hep-th/0007191}{{\tt
  hep-th/0007191}}].

\bibitem{Aharony:2007vg}
O.~Aharony, A.~Buchel and P.~Kerner, \emph{{The Black hole in the throat:
  Thermodynamics of strongly coupled cascading gauge theories}},
  \href{http://dx.doi.org/10.1103/PhysRevD.76.086005}{\emph{Phys. Rev.} {\bf
  D76} (2007) 086005}, [\href{https://arxiv.org/abs/0706.1768}{{\tt
  0706.1768}}].

\bibitem{Buchel:2005nt}
A.~Buchel, \emph{{A Holographic perspective on Gubser-Mitra conjecture}},
  \href{http://dx.doi.org/10.1016/j.nuclphysb.2005.10.014}{\emph{Nucl. Phys.}
  {\bf B731} (2005) 109--124},
  [\href{https://arxiv.org/abs/hep-th/0507275}{{\tt hep-th/0507275}}].

\end{thebibliography}\endgroup

\end{document}